\documentclass[conference]{IEEEtran}
\usepackage[utf8]{inputenc}
\usepackage[T1]{fontenc}
\usepackage[linesnumbered,ruled,vlined]{algorithm2e}
\usepackage{subcaption}
\usepackage{float}
\usepackage{multirow}
\usepackage{booktabs}
\usepackage{bm}
\usepackage{graphicx}
\usepackage{balance}

\usepackage{array}

\ifCLASSINFOpdf
\else
\fi
\begin{document}
%
\title{Urban Crowdsensing using Social Media: An Empirical Study on Transformer and Recurrent Neural Networks}

\author{\IEEEauthorblockN{Jerome Heng\IEEEauthorrefmark{1}, Junhua Liu\IEEEauthorrefmark{1} and Kwan Hui Lim\IEEEauthorrefmark{1}}
\IEEEauthorblockA{\IEEEauthorrefmark{1}Information Systems Technology and Design Pillar, Singapore University of Technology and Design\\}
\IEEEauthorblockA{Email: \{jerome\_heng, junhua\_liu\}@mymail.sutd.edu.sg,  kwanhui\_lim@sutd.edu.sg}
}


%


\IEEEoverridecommandlockouts
\IEEEpubid{\makebox[\columnwidth]{978-1-7281-6251-5/20/\$31.00~\copyright~2020 IEEE \hfill} \hspace{\columnsep}\makebox[\columnwidth]{ }}

\maketitle

\IEEEpubidadjcol

\begin{abstract}
An important aspect of urban planning is understanding crowd levels at various locations, which typically require the use of physical sensors. Such sensors are potentially costly and time consuming to implement on a large scale. To address this issue, we utilize publicly available social media datasets and use them as the basis for two urban sensing problems, namely event detection and crowd level prediction. One main contribution of this work is our collected dataset from Twitter and Flickr, alongside ground truth events. We demonstrate the usefulness of this dataset with two preliminary supervised learning approaches: firstly, a series of neural network models to determine if a social media post is related to an event and secondly a regression model using social media post counts to predict actual crowd levels. We discuss preliminary results from these tasks and highlight some challenges. 
\end{abstract}


%
\IEEEpeerreviewmaketitle

\section{Introduction}

With increasing urbanization in the world, it is important for urban planners to understand crowd levels for better urban planning~\cite{al2013crowd}. Many users have made their social media posts publicly accessible and this can potentially be a source of useful insights for urban analytics, such as identifying events, detecting crowd levels and other environmental phenomenon~\cite{zappatore2017crowd,Yong-BigData20}. While another possible solution is to employ the use of physical sensors, such sensors are potentially costly and time consuming to implement on a large scale.

The challenge with using social media is that it is often unstructured and sparse, hence great care must be taken to filter the dataset such that the information is relevant, and aggregation functions can be used in appropriate places to reduce sparsity~\cite{ranneries2016wisdom}. Existing solutions are based on social media filtering and clustering techniques, e.g., sending the data through a pipeline to detect clusters. Researchers at Twitter have constructed a real-time pipeline to detect clusters of trending tweets and topics~\cite{fedoryszak2019real}. Above a certain threshold, these clusters are identified as events/outliers. 

We intend to extend and improve on such approaches for urban crowdsensing using social media. In this paper, we make the following contributions:
\begin{itemize}
    \item We curate and introduce two datasets that can be used for various urban sensing applications, particularly for detecting crowd levels and identifying urban events with the corresponding ground truth labels (Section~\ref{sectDataset}). In developing these datasets, we also identify the limitations of social media data and the steps we took to overcome these limitations. This dataset is publicly available at https://github.com/kraftedcheese/crowdsensing\_datasets.
    \item We proposed two problems that can be investigated on this dataset. The first problem is an event detection problem where we determine if a specific social media post is related to an urban event that is characterized by higher human traffic at a given location and time. The second problem is a crowd level prediction problem that is modelled as a regression problem where we predict the relative counts of people in various areas of a city from social media posts in that area.
    \item For our first problem on event detection, we experiment with a set of transformer and recurrent neural networks models and report some preliminary results. For the second problem on crowd level prediction, we utilize a regression model using social media post counts to predict actual crowd levels.
\end{itemize}

\section{Dataset}
\label{sectDataset}
Geo-tagged social media has been frequently used for solving various location-based problems, such as predicting the location of a social media posts~\cite{chong2019fine}, predicting next visit locations~\cite{altaf2018spatio} and recommending tour itineraries~\cite{Liu-ECMLPKDD20}, among others. Similarly, we curate our dataset from mainly geo-tagged Twitter messages (i.e., tweets) and Flickr photos, which we retrieve using the Twitter and Flickr APIs, respectively. Tables~\ref{tab:dataEvent} and~\ref{tab:dataCrowd} gives an outline of our curated dataset of Twitter geo-tagged messages and Flickr geo-tagged photos. 

In the following sections, we will describe them in more details as we demonstrate their application for an event detection task and crowd prediction task.

\begin{table}[h]
\renewcommand*{\arraystretch}{1.2}
\centering
{%
\begin{tabular}{lc}
\hline\hline
Date	&	2014-01-12 to 2018-02-12    \\	
Number of Tweets    &	1,393,561   \\	
Number of Unique Users  &	77,642  \\	
Number of Events    &   22,241 \\
Number of Non-events    &   1,371,320  \\ \hline\hline
\end{tabular}}
\caption{Dataset for Event Detection}
\label{tab:dataEvent}
\end{table}

\begin{table*}[t]
\renewcommand*{\arraystretch}{1.25}
\centering
{%
\begin{tabular}{lcccc}
\hline \hline
	&		&	Number of	&	Number of	&	Number of \\
Name	&	Date	&	Data Points	&	crowd sensors	&	unique users \\ \hline
Sensor Counts 	&	2009-05-01 to 2020-04-30	&	3,132,346	&	66	&	-	 \\
Tweets (Sensor-related)	&	2010-09-12 to 2018-06-04	&	266,931	&	-	&	20,176	 \\
Tweets (max 100 duplicate coords.)	&	2010-09-21 to 2018-02-02	&	24,638	&	42	&	5,659 \\
Tweets (max 10 duplicate coords.)	&	2010-09-21 to 2018-02-02	&	12,801	&	42	&	3,785 \\
Flickr (near to Town Hall (West))	&	2010-07-08 to 2020-03-29	&	6,076	&	-	&	852 \\ \hline
\hline
\end{tabular}}
\caption{Dataset for Crowd Level Prediction}
\vspace{1mm}
\label{tab:dataCrowd}
\end{table*}

\begin{table*}[th]
\renewcommand*{\arraystretch}{1.25}
\centering
{%
\begin{tabular}{lccccccccc}
\hline \hline
 &   \multicolumn{3}{c}{Weighted Average}   &   \multicolumn{3}{c}{"Event" Class}   &   \multicolumn{3}{c}{"Non-Event" Class}\\
Model   &   Precision   &   Recall    & F1   &   Precision   &   Recall    & F1   &   Precision   &   Recall    & F1\\ \hline
LSTM & 0.7080 & 0.7292 & 0.6863 & 0.6267 & 0.2489 & 0.3563 & 0.7431 & 0.9361 & 0.8285\\ 
GRU & 0.7230 & 0.7205 & 0.6400 & 0.7299 & 0.1141 & 0.1974 & 0.7200 & 0.9818 & 0.8308 \\
S-LSTM & 0.7130 & 0.7078 & 0.6042 & 0.7260 & 0.0476 & 0.0894 & 0.7074 & 0.9923 & 0.8260 \\
DistilBERT & 0.4884  &  0.6989  &  0.5750 & 0.0000  &  0.0000  &  0.0000 & 0.6989  &  1.0000  &  0.8227 \\ 
\hline \hline
\end{tabular}}
\caption{Results for the Event Detection Task}
\label{tab:resultsEvent}
\end{table*}

\subsection{Events Dataset}

Table~\ref{tab:dataEvent} shows the summary statistics of our event detection dataset. This dataset comprises a set of tweets and the corresponding list of events. As a proxy for events label, we utilize a list of all events in Melbourne that had an event permit issued\footnote{https://data.melbourne.vic.gov.au/Events/Event-permits-2014-2018-including-film-shoots-phot/sex6-6426}. As a further pre-processing, we filtered out events that were likely to not be public, such as film/movie shoots. Using the Twitter API, we then extracted the tweets posted during each event and within 100m of the event location, as well as all tweets in Melbourne from the first event date to the last event date. Following which, this resulted in a dataset of 1.39 million tweets, of which 22k were tweets related to an event

\subsection{Crowd Level Dataset}

Table~\ref{tab:dataCrowd} summarizes the main statistics of our crowd level prediction dataset. There are two components to this dataset, namely a list of pedestrian sensors and their pedestrian counts and the geo-tagged tweets in the vicinity of these sensors. The pedestrian sensor component comprises 3.1 million pedestrian count readings from 66 sensors, retrieved from Melbourne Open Data\footnote{https://data.melbourne.vic.gov.au/Transport/Pedestrian-Counting-System-2009-to-Present-counts-/b2ak-trbp}. In addition, we collected 266k geo-tagged tweets posted within Melbourne City between 2010 to 2018. These tweets were then mapped to individual pedestrian sensor locations if their distances differ by less than 100m.

Upon inspecting the resultant dataset, we realized that many of the tweets had duplicate coordinates. 
A likely cause of this issue is the tendency for users to tag tweets with Twitter “places”~\cite{botta2015quantifying}, which are polygonal areas corresponding to places of interest that have a standard set of coordinates; in contrast to having their phone GPS report their accurate coordinates. To address this issue, we wrote a script to check for duplicate lat/long coordinates. We set a threshold of 100 maximum duplicates, which left us with a filtered dataset 10\% of its original size.

\section{Event Detection}

For the event detection task, one concern is with the imbalance distribution of classes. To address this issue, we first performed undersampling as our dataset is very imbalanced with a minority of non-events class~\cite{fernandez2018learning}. After this undersampling process, we obtain a dataset with approximately 30\% events and 70\% non-events, comprising a total of 74,136 tweets. Using this set of tweets, we used 90\% for training and 5\% each for our development and test set.

In our experiments, we empirically evaluate the following neural network based models, namely: Long Short-Term Memory (LSTM)~\cite{hochreiter1997long}, Gated Recurrent Unit (GRU)~\cite{cho2014learning}, Stacked LSTM (S-LSTM) and DistilBERT~\cite{sanh2019distilbert}. In these recurrent neural networks, we used GloVe embedding~\cite{pennington2014glove} to represent the individual words in the tweets. We also empirically select the best performing values for the learning rate, dropout rate, hidden states, batch size on the development set and report the performance on the test set.

\begin{figure*}[th]
    \centering
    \includegraphics[width=0.49\textwidth]{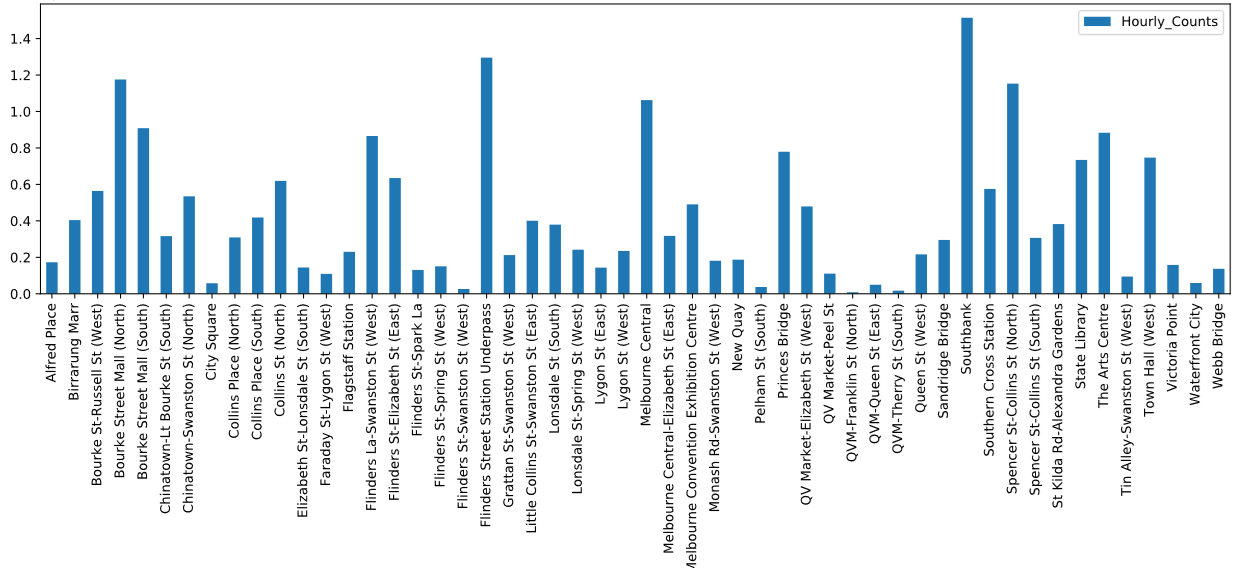}
    \includegraphics[width=0.49\textwidth]{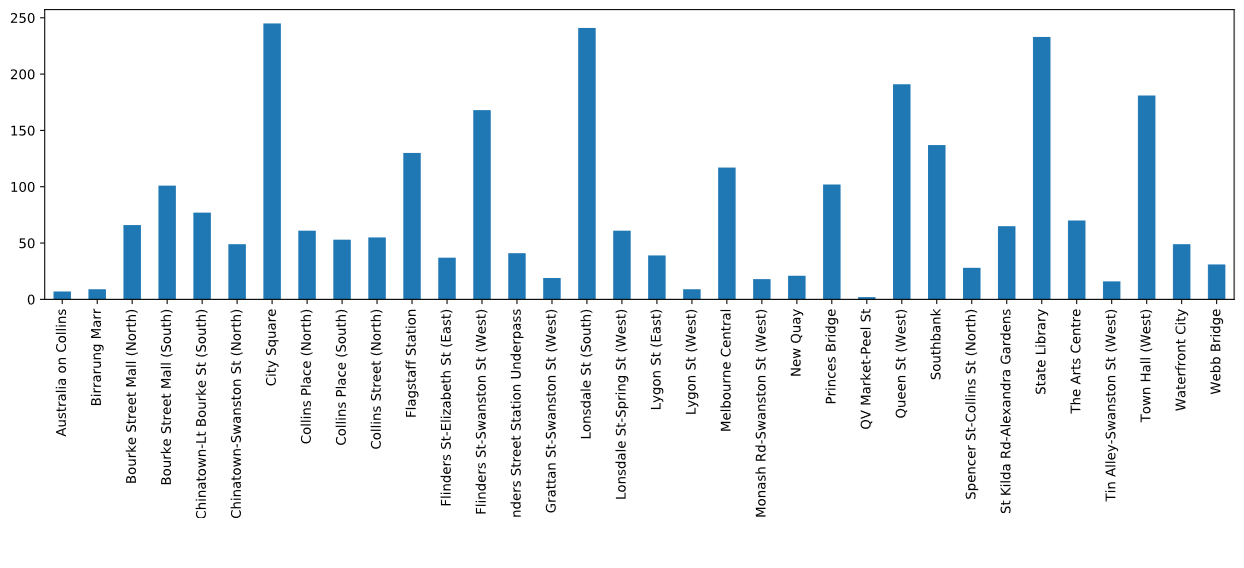}
    \caption{Frequency counts of sensor-based pedestrian counts (left) and tweet counts by sensor locations (right)}
    \label{fig:sensorTweetCount}
\end{figure*}

For our evaluation, we use the commonly used metrics of precision, recall and F1-score. Table~\ref{tab:resultsEvent} shows the results of our preliminary experiments in terms of precision, recall and F1-score for each class as well as the weighted mean. 
In terms of Weighted F1-score, LSTM is the best performer, followed by GRU, S-LSTM and DistilBERT. We believe that the smaller training dataset after the undersampling process contributed to a poorer performance, particularly for DistilBERT. Moving forward, we intend to further augment this dataset with more tweets that correspond to the "Event" class and repeat the experiments on similar datasets.

\section{Crowd Level Exploratory Analytics}
By plotting the tweet and pedestrian count by sensor location in a specific year, we observe that there is some correlation between the two; where there are high pedestrian counts near a particular sensor location, the corresponding number of tweets is likely to also be high. An example is shown in Figure~\ref{fig:sensorTweetCount}. The exceptions to this trend are largely due to some sensors not having any linked tweets, especially when there are many sensors close together (e.g in the city center). 

These filtering algorithms left us with a very sparse twitter dataset compared to our pedestrian count dataset. For instance, taking the counting sensor at Town Hall (West) in 2017, we had a tweet range from 0-500 per hour and a pedestrian count range from 200-1000 per hour. Based on a scatter plot of the tweets and pedestrian hourly counts, we observed a large proportion of data points where there were no tweets and a varying number of pedestrian counts. This makes it challenging to find meaningful correlations of the tweet counts to the pedestrian counts from this scatter plot. This was confirmed by an initial pass through of the data to a few logistic regression and linear regression algorithms which showed little to any usable results – the average accuracy was below 10\% for the test set. 

A dataset used for crowd level prediction must have reduced sparsity, for instance by including data from other sources. Flickr is one such source that we looked at. Flickr is particularly suitable for our purposes, because geo-location data in Flickr comes from the camera’s metadata. We used the Flickr API to crawl posts from the same timeframe as the Twitter dataset. Thus, we can use Flickr descriptions to augment the text data from tweets. However, a detailed method of extracting meaningful data patterns from a combined data source requires more investigation.

\section{Conclusion}
Our main contribution is in curating two datasets, one for event detection and one for crowd level prediction. We identify the difficulties of creating a dataset for crowd level regression – particularly the issues relating to data sparsity and geo-location accuracy. We also discussed some preliminary results on the two problems of event detection and crowd level prediction. For the event detection problem, we experimented with an initial set of neural network models, such as LSTM, GRU, Staked LSTM and DistilBERT, and discuss preliminary results and their limitations. For the crowd level prediction problem, we performed exploratory data analytics to identify trends between social media data and pedestrian sensors.

\section{Acknowledgement}

This research is funded in part by the Singapore University of Technology and Design under grant SRG-ISTD-2018-140.



\bibliographystyle{IEEEtran}
\balance
\bibliography{ref}
%



\end{document}